\documentclass[letters,usenatbib]{mnras}

\usepackage{newtxtext,newtxmath}

\usepackage[T1]{fontenc}

\DeclareRobustCommand{\VAN}[3]{#2}
\let\VANthebibliography\thebibliography
\def\thebibliography{\DeclareRobustCommand{\VAN}[3]{##3}\VANthebibliography}

\usepackage{graphicx}	
\usepackage{amsmath}	

\title[Centrifugal force and retrievals]{Effect of Centrifugal Force on Transmission Spectroscopy of Exoplanet Atmospheres}

\author[A. Banerjee et al.]{
Agnibha Banerjee,$^{1}$\thanks{E-mail: agnibha.banerjee@open.ac.uk}
Joanna K. Barstow,$^{1}$
Carole A. Haswell$^{1}$
and Stephen R. Lewis$^{1}$
\\
$^{1}$School of Physical Sciences, The Open University, Milton Keynes, MK7 6AA, UK
}

\date{Accepted XXX. Received YYY; in original form ZZZ}

\pubyear{2023}

\begin{document}
\label{firstpage}
\pagerange{\pageref{firstpage}--\pageref{lastpage}}
\maketitle

\begin{abstract}
Transmission spectroscopy is one of the most successful methods of learning about exoplanet atmospheres. The process of retrievals using transmission spectroscopy consists of creating numerous forward models and comparing them to observations to solve the inverse problem of constraining the atmospheric properties of exoplanets. We explore the impact of one simplifying assumption commonly employed by forward models of transiting exoplanets: namely that the planet can be treated as an isolated, non-rotating spherical body. 
The centrifugal acceleration due to a planet's rotation opposes the gravitational pull on a planet's atmosphere and increases its scale height. Conventional forward models used for retrievals generally do not include this effect. We find that atmospheric retrievals produce significantly different results for close-in planets with low gravity when this assumption is removed, e.g.,
differences between true and retrieved values of gas abundances greater than 1$\sigma$ for a simulated planet analogous to WASP-19 b. We recommend that the correction to the atmospheric scale height due to this effect be taken into account for the analysis of high precision transmission spectra of exoplanets in the future, most immediately JWST Cycle 1 targets WASP-19\,b and WASP-121\,b.

\end{abstract}




\section{Introduction}

The study of exoplanet atmospheres through transmission spectroscopy using retrievals is now a well-established method to deduce their composition and thermal structure \citep{char02, Tinetti2007, sing2011a, Kreidberg2014}. Robust detections of atmospheric species such as $\textrm{Na}$, $\textrm{K}$, $\textrm{H}_2\textrm{O}$ \citep{wakeford2022} have already been made with previous studies using space-based telescopes such as the Hubble Space Telescope and the Spitzer Space Telescope. Evidence of some carbon-bearing species such as $\textrm{CO}_2$ \citep{alderson2018} have also been found. Recently, definitive evidence has been found to confirm the presence of $\textrm{CO}_2$ in the atmosphere of WASP-39b \citep{ersco2} using JWST. With the increase in quality of spectral data owing to JWST, the analysis of much finer aspects of these atmospheres is now possible. This jump in signal-to-noise ratio will warrant the removal of assumptions that have so far been considered routine. In this letter, we explore one such assumption and evaluate the degree to which it impacts our inferences regarding atmospheric properties.

When a planet spins, the centrifugal force due to rotation acts to oppose the gravitational force, resulting in a slight reduction in the effective gravitational acceleration acting on the atmosphere. The gravitational equipotentials and hence the effective gravity in a two body system like a planet orbiting a star has been well-studied in the context of close binary stars, see e.g. \cite{frank_king_raine_2002}, and this Roche geometry has been applied to exoplanets in \cite{2017PhDT.......326B} and \cite{oblate}. Atmospheric scale height is defined as the altitude at any point in the atmosphere over which the atmospheric pressure reduces by a factor of $e$, and is a quantity often used to measure the extent of an atmosphere. The scale height is inversely proportional to effective gravitational acceleration and thus increases when the centrifugal force is included in the effective gravity calculation. The modified scale height, in turn, affects the transmission spectra obtained from it by increasing the observable feature amplitudes. Whilst the effect of rotation on the effective gravity is well known and commonly accounted for in models of the fast-rotating Solar System giant planets \citep{lindal1985, patbook2009}, it has not generally been included in retrieval models for exoplanets.

In this letter, we analyse the impact of reduced effective gravity on synthetic transmission spectra for a range of hypothetical planets and also examine the resulting influence on retrieved atmospheric properties. We first describe the atmospheric retrieval code, NEMESIS (Non-linear optimal Estimator for MultivariatE spectral analySIS) \citep{Irwin2008}. We then demonstrate the difference in transmission spectra with and without a latitudinally-averaged correction due to centrifugal forces, and perform two retrievals on the transmission spectrum generated with the correction included: one with an atmospheric model that includes the correction, and one without. We find that the retrieval model without the centrifugal correction is unable to recover the input atmospheric state to within 1$\sigma$. Finally, we perform a parameter space exploration of planetary bulk density, equilibrium temperature, stellar radius and orbital period to constrain the regions in which this correction creates a significant difference in the retrieval results.

\section{Methods}
\subsection{The NEMESIS Code}
We use the NEMESIS algorithm to create synthetic spectra and also to generate forward models for the retrievals. Originally developed for the characterisation of the atmospheres of Solar System planets, NEMESIS has since been modified for use in studying exoplanet atmospheres \citep{Lee2012}. NEMESIS uses a correlated-k radiative transfer model \citep{lacis1991JGR....96.9027L}. We use the nested sampling algorithm option \citep{krissansen18} to perform retrievals as it allows the exploration of non-Gaussian posterior distributions.
We use the pymultinest package \citep{Buchner2014} based on the MultiNest \citep{Skilling2004,feroz09} algorithm, which is widely used for retrieval of exoplanet transmission spectra.

\begin{figure}
	\includegraphics[width=\columnwidth]{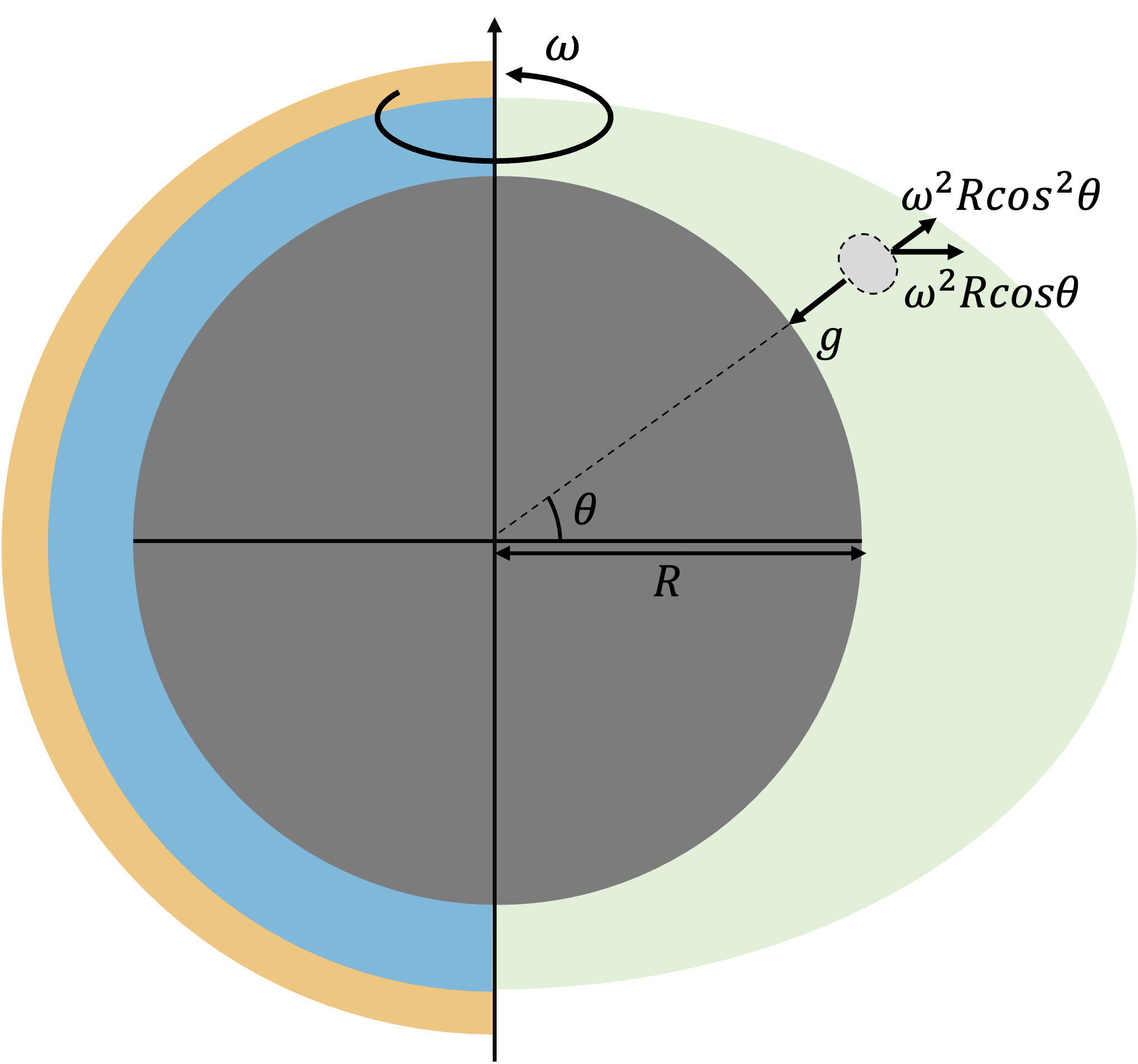}
    \caption{A schematic cross section of a planet along the plane of the terminator showing the accelerations acting on an atmospheric parcel (light grey). On the left half of the diagram, the atmospheric extent corresponding to the scenario with rotation included is illustrated by an orange annulus (outer). The atmospheric scale heights in the diagram have been exaggerated for visibility. The atmospheric extent corresponding to the scenario without the rotation included is illustrated by a blue annulus (inner). On the right half of the diagram, an exaggerated elliptical atmospheric annulus is shown. The centrifugal acceleration acting on the parcel is perpendicular to the axis of rotation and marked as $\omega^2 R \cos \theta$, and the radial component of this acceleration which opposes the gravitational acceleration $g$ is marked as $\omega^2 R \cos^2 \theta$. In the quantitative parts of this paper, we approximate the atmospheric annulus to be circular.}
    \label{fig:scheme}
\end{figure}

\subsection{Analytic Estimate of the Magnitude of the Centrifugal Effect}
\begin{figure}
	\includegraphics[width=\columnwidth]{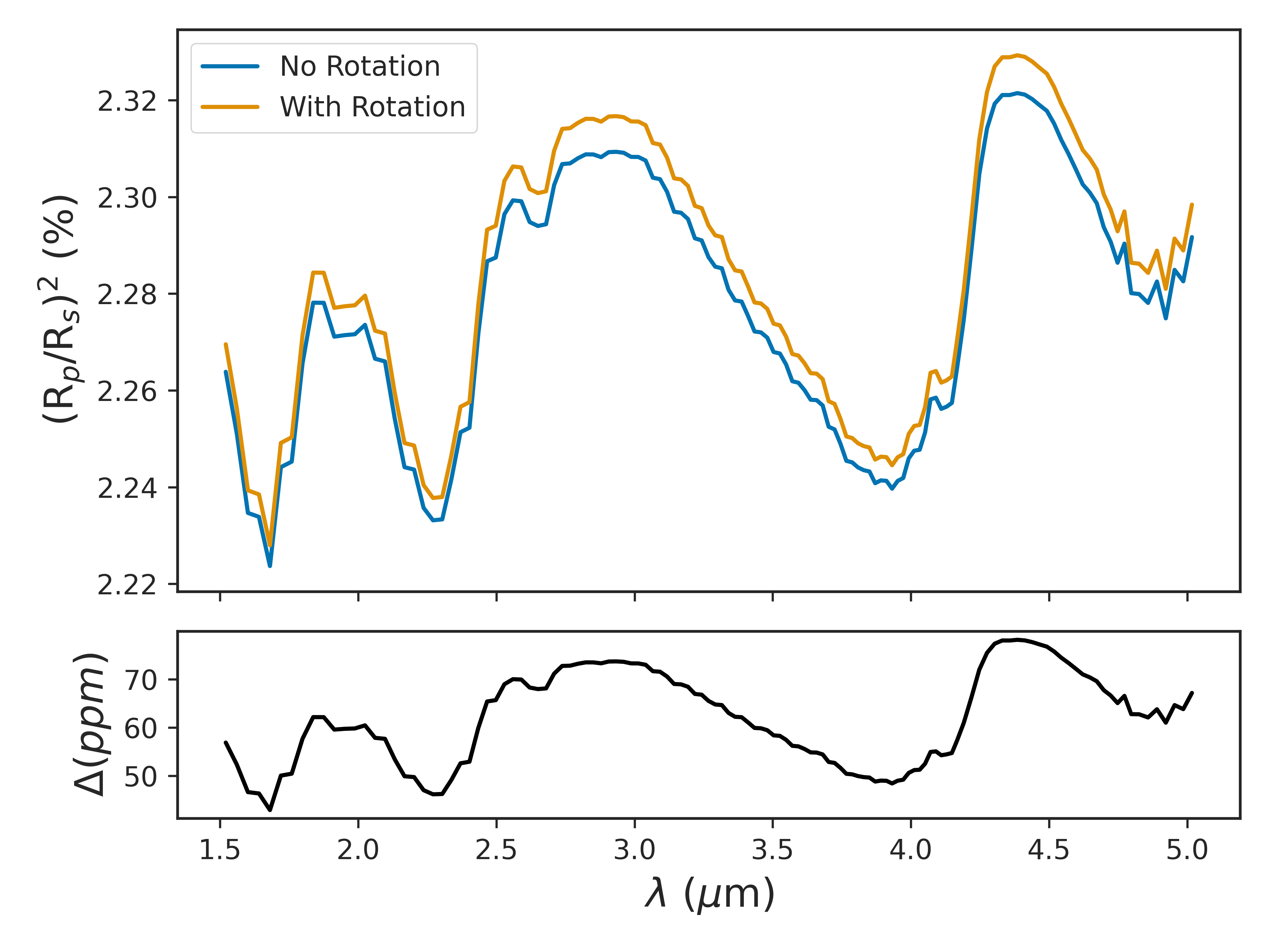}
    \caption{The difference between the simulated spectra with (orange) and without (blue) the effect of centrifugal acceleration included. The planet used is a WASP-19 b analogue, with parameters as described in Section \ref{secmod}. The forward model with this effect taken into consideration has a more extended atmosphere and, thus, a higher value for the observed planetary radius at each wavelength. The increase in transit depth is a scaled version of the base spectra and, thus, is wavelength dependent.} 
    \label{fig:rotaspec}
\end{figure}

As an initial assessment of the magnitude of the change in the transmission spectrum due to the scale height being modified by rotation,
a synthetic spectrum was created using NEMESIS with the modified value of the acceleration due to gravity. Figure \ref{fig:rotaspec} demonstrates the difference between the simulated spectra with and without the effect of centrifugal acceleration included. The atmospheric model used is described in detail in Section \ref{secmod}.

In transmission spectroscopy, we only sample the annulus of the atmosphere around the terminator of an exoplanet at superior conjunction of the planet, i.e. during planet transit. Though some close-in planets which are particularly attractive targets for transmission spectroscopy are significantly non-spherical \citep[see e.g. Fig. 1 of][]{staab2017}, the cross-section viewed during transit has minimal deviation from a circular cross-section. Thus we can treat the annulus as being bounded by concentric circles.
We discuss the implications of non-sphericity further in Section~\ref{sec:oblate}. We also assume that the planet rotates as a rigid body. 


The acceleration caused by the apparent outward force due to rotation is perpendicular to the axis of rotation and is computed as $\omega^2 R \cos \theta$, where $\omega$ is the angular velocity, $R$ is the radius of the planet, and $\theta$ is the latitude (Figure~\ref{fig:scheme}). The radially outwards component of this acceleration is, therefore $\omega^2 R \cos^2 \theta$. We assume that the planet is tidally locked, and thus the orbital period is the same as the rotation period, and the axis of rotation of the planet is perpendicular to the orbital plane \citep{heller2011}.

An average value of net gravitational acceleration over the annulus relevant to transmission spectroscopy is then obtained by integrating over latitudes from the equator to the pole and then dividing by the range of latitudes covered. 

\begin{equation*}
\begin{split}
g'(\theta,R') &= g(R')-\omega^{2} R' \cos ^{2} \theta \\
g_{\rm{av}}'(R') &= \frac{\int_{0}^{\frac{\pi}{2}} g'(\theta,R') \textrm{d} \theta}{\int_{0}^{\frac{\pi}{2}} \textrm{d} \theta}\\
g_{\rm{av}}'(R') &= \frac{\int_{0}^{\frac{\pi}{2}}\left(g(R')-\omega^{2} R' \cos ^{2} \theta\right) \textrm{d} \theta}{\frac{\pi}{2}} \\
&= g(R')-\frac{2 \omega^{2} R'}{\pi} \int_{0}^{\frac{\pi}{2}} \cos ^{2} \theta \textrm{d} \theta \\
&= g(R')-\frac{1}{2} \omega^{2} R'
\end{split}
\label{eq:lat_avg}
\end{equation*}

This corrected value of $g$ impacts the calculated atmospheric scale height in the forward model. Scale height $H$ is given by
\begin{equation*}
H = \frac{kT}{\mu{ } g_{\rm{av}}'(R')}
\end{equation*}
where $k$ is the Boltzmann constant, $T$ is the atmospheric temperature and $\mu$ is the mean molecular weight of the atmosphere. Since the scale height varies as the inverse of gravitational acceleration, it is increased when the modified value of gravitational acceleration is used. The extent of atmospheric features in the transmission spectrum is proportional to $H$, so the increase in $H$ produces a stretch in the transmission spectrum (Figure~\ref{fig:rotaspec}).

\subsection{Retrieval Set-Up}
\label{secmod}
The NEMESIS model used assumes a cloud-free, isothermal, well-mixed atmosphere containing $\textrm{H}_2\textrm{O}$ and $\textrm{CO}_2$ as the spectrally active trace gases and a background made up of $\textrm{H}_2$ and $\textrm{He}$ ($\textrm{He}$:$\textrm{H}_2$ = 0.17). The $\textrm{H}_2\textrm{O}$ k-table was obtained from \citet{exomol_h2o} and the $\textrm{CO}_2$ k-table was obtained from \citet{exomol_co2}, both tables generated as in \cite{katy2021A&A...646A..21C} Collision-induced absorption from H$_2$ and He is from \cite{Borysow_2002}. A simplistic atmospheric model was adopted as we wish to isolate the effect of rotation from all other complexities. The model goes to a maximum pressure of 10 bar, and the planet radius is defined as the radius at 100 mbar. Priors and ranges for each parameter are included in Table~\ref{table:params}.

The planet used is an analogue of WASP-19 b, with a radius of 1.415 $\textrm{R}_{\textrm{Jup}}$, a mass of 1.154 $\textrm{M}_{\textrm{Jup}}$, equilibrium temperature of 2113.0 K and an orbital period of 0.78 days \citep{wasp19b_params}. It is used to demonstrate the impact of rotation on retrievals as it has a large atmospheric scale height. An error envelope of 30 ppm was used over the simulated spectrum which included the modification in $g$ due to rotation. A spectral range of 1.5 $\mu$m -- 5.0 $\mu$m was used, with a spectral resolution ($\lambda / \Delta \lambda$) ranging between 60-200 -- corresponding to the binned resolution for the JWST NIRSpec PRISM instrument used in \citet{ersco2}. No additional scatter noise was added to the simulated spectrum.

\begin{table*}
\begin{tabular}{|l|l|l|l|l|l|}
\hline
Parameter   & Prior Type       & Prior Range    & True Value & Retrieved Value (With Spin) & Retrieved Value (No Spin) \\ \hline
$\log(\textrm{H}_2\textrm{O})$        & log-uniform & -12, -1        & -2.00      & $-2.020^{+0.054}_{-0.061}$                     & $-2.373^{+0.129}_{-0.229}$                       \\
$\log(\textrm{CO}_2)$        & log-uniform & -12, -1        & -4.00      & $-4.024^{+0.065}_{-0.071}$                     & $-4.380^{+0.142}_{-0.246}$                       \\
$\textrm{T}_{\textrm{iso}}$ (K)      & uniform & 1600.0, 2400.0 & 2113.00    & $2108.95^{+27.55}_{-27.91}$                   & $2108.35^{+27.51}_{-28.21}$                     \\
Radius ($\textrm{R}_{\textrm{Jup}}$) & uniform & 1.1, 1.8       & 1.415        & $1.415^{+0.001}_{-0.001}$                     & $1.421^{+0.003}_{-0.002}$                       \\ \hline
\end{tabular}
\caption{The true values used for the simulated spectrum and the retrieved values for the cases with and without planetary rotation, obtained from the sample retrieval. The values reported are the median and the $1 \sigma$ errors.}
\label{table:params}

\end{table*}
\section{Results}

\subsection{Retrieval}
\label{sec:retrieve}
Two retrievals were performed on the synthetic spectrum including centrifugal effects which is shown in orange in Fig.~\ref{fig:rotaspec};
one including centrifugal effects in the retrieval model, and one assuming zero rotation. Table \ref{table:params} compares the retrieved values with the inputs for each parameter.

Both retrievals produced an equally good fit to the input spectrum, but the retrieved parameters varied between them, and the retrieval excluding rotational effects does not correctly recover the input atmospheric state. The retrieved gas abundances are $\sim 0.3 \textrm{dex}$ (or $\sim 3\sigma$) lower than their true values when the retrieval is performed without the corrections for spin. The posterior distributions for the retrieval excluding spin also have increased skewness. In the scenario without rotation included, a higher value is retrieved for the radius of the planet which compensates for the increased transit depth, actually caused by
lower gravity and higher scale height due to rotation. However, increasing the radius lifts up the spectrum while also stretching feature amplitudes. 
The gas abundances in this scenario are retrieved to be lower than the truth to compensate for this effect and reduce the feature amplitudes to match the simulated spectrum. This test indicates that retrievals that do not include the centrifugal effect can be apparently successful, whilst producing spurious results.

\begin{figure}
    \includegraphics[width=\columnwidth]{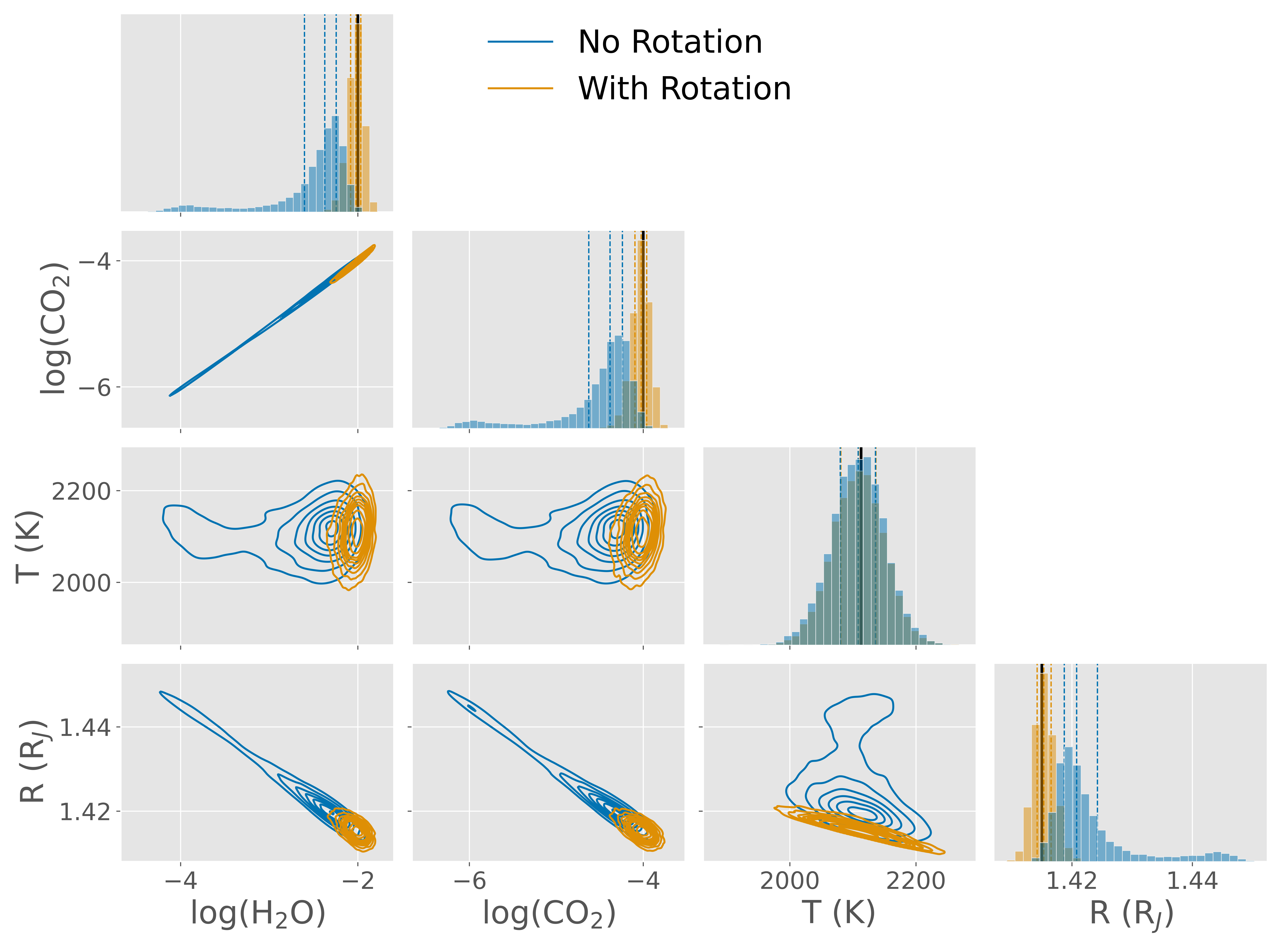}
    \caption{A corner plot showing the difference in retrieved parameters with (orange) and without (blue) the correction to $g$ due to planetary rotation included. A significant shift can be seen in the retrieved gas abundances and planetary radius.}
    \label{fig:retrieval}
\end{figure}

\subsection{Parameter Space Exploration}
We used a grid of planetary parameters to test which planets would be most affected by the inclusion of the modified effective gravity in the calculation of the scale height. The planetary parameters varied were the planet's radius, equilibrium temperature, stellar radius and orbital period. The planetary radius was varied from 1.0 to 2.0 $\textrm{R}_{\textrm{Jup}}$, the equilibrium temperature was varied from 1000 K to 2000 K, the stellar radius was varied from 0.8 $\textrm{R}_{\sun}$ to 1.5 $\textrm{R}_{\sun}$, and the rotation period was varied from 0.6 days to 4.5 days. The planetary mass was set to 0.8 $\textrm{M}_{\textrm{Jup}}$ \footnote{$\textrm{R}_{\textrm{Jup}}$ = 69911 km, $\textrm{M}_{\textrm{Jup}}$ = 1.898 $\times 10^{27}$ kg, $\textrm{R}_{\sun}$ = 695700 km}. All other parameters used in the atmospheric forward model setup were kept the same, as described in Section~\ref{secmod}.
The equilibrium temperature divided by bulk density and the square of the stellar radius (Similar to the Transmission Spectroscopy Metric, see \citet{tsm})
was used as a metric for how inflated the planet's atmosphere is. Denser planets have a greater gravitational pull and combined with lower equilibrium temperatures, they are expected to have lower atmospheric scale heights, and vice versa. A greater value of stellar radius also contributes to a lower relative transit depth. 

\begin{figure*}
    \includegraphics[width=\textwidth]{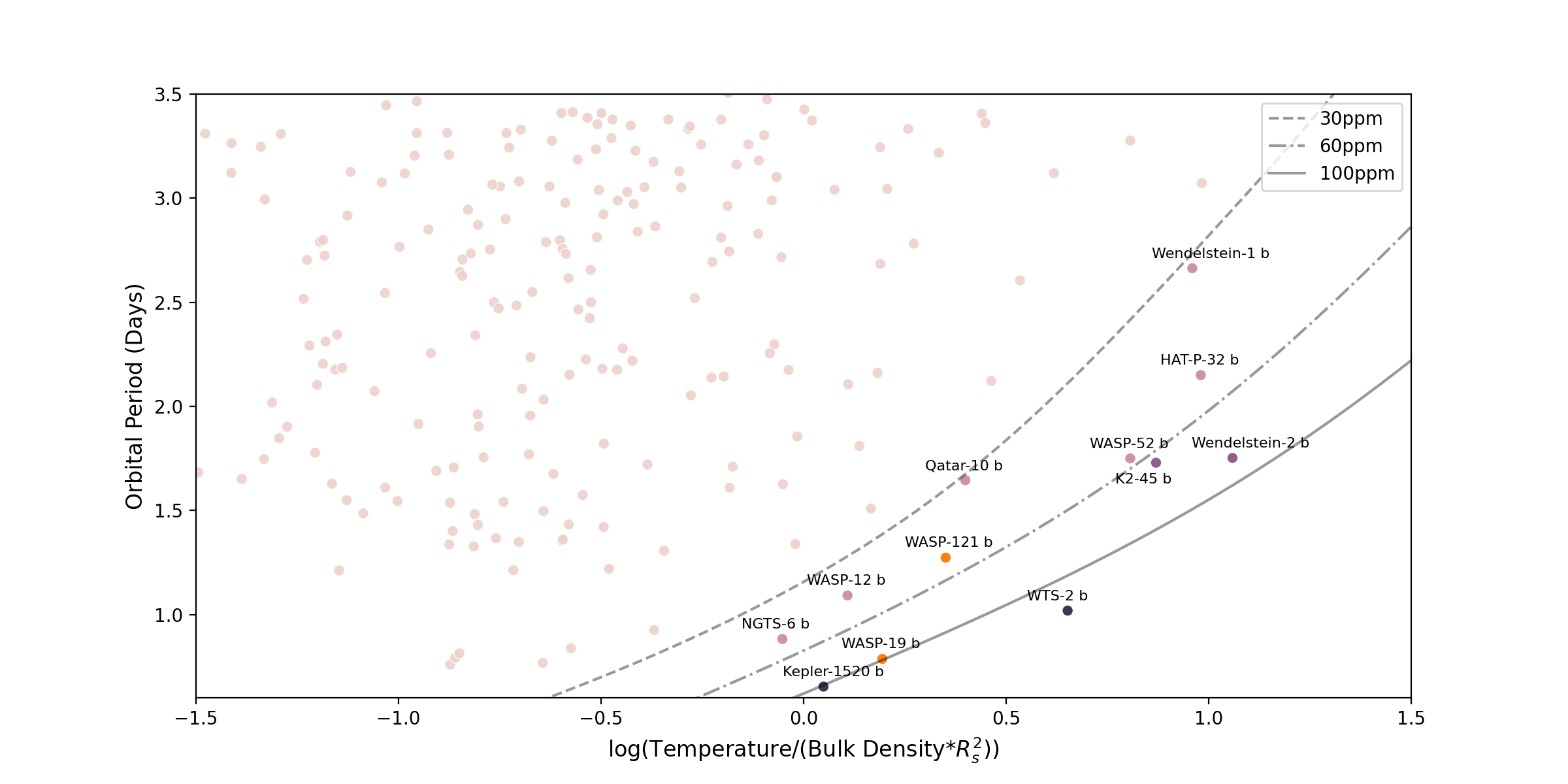}
    \caption{Orbital period versus a metric of planet atmosphere inflation with real planets plotted. Lines indicate the boundaries where the offset between transmission spectra with and without rotation reach values of 30\,ppm, 60\,ppm, and 100\,ppm. The change in the transmission spectrum due to rotation is strongest towards the lower right. 
    Where the effect of rotation exceeds the 30 ppm limit, the planets are marked with their names.
    Cycle 1 JWST targets WASP-19 b and WASP-121 b are also marked and highlighted in orange.}
    \label{fig:boundary}
\end{figure*}

In Figure \ref{fig:boundary}, we illustrate the magnitude of the centrifugal correction over a range of planetary parameters. The synthetic spectra with and without rotation were compared at each grid point, and the average difference between them was calculated. Bounding curves to denote values of this difference at 30, 60 and 100 ppm \citep{Barstow2017, Taylor2022} respectively, are plotted. We can distinguish the change in the transmission spectrum due to rotation from random noise at either 30, 60 or 100 ppm, for the planets that lie below and to the right of each bounding curve.

Real planets are overplotted on this figure, with the planetary radius, mass, equilibrium temperature and host star radius for each obtained from the NASA Exoplanet Archive. Two of the exoplanets which would show significant differences in retrieved parameters with rotation included are Cycle 1 JWST targets: WASP-19 b and WASP-121 b. The exact magnitude of the effect will depend not only on the bulk properties of the planet but also on its atmospheric composition, and therefore the location of a particular planet in Figure~\ref{fig:boundary} is intended as a guide rather than an exact statement of how large the effect will be in each case.

\section{Discussion}
\subsection{Dependence of Centrifugal Acceleration on Latitude and Planetary Oblateness}
\label{sec:oblate}
We have not accounted for the variation of centrifugal acceleration with latitude in this process. To demonstrate that taking the latitudinal average is a reasonable approximation, we generated a transmission spectrum for a planet segmented into 20 latitudinal sections. The effective gravitational acceleration for each segment increases towards both poles as the perpendicular distance from the rotation axis decreases. However, the difference between this transmission spectrum and one generated without considering the latitudinal variation for realistic values of the orbital period (1 day), planetary mass (1 $\textrm{M}_{\textrm{Jup}}$), planetary radius (1 $\textrm{R}_{\textrm{Jup}}$), and equilibrium temperature (2000 K) is of the order of 5 ppm.

We have also not accounted for the oblateness of a planet. Recent works \citep{oblate,2017PhDT.......326B} have considered the effect of the oblateness of a planet on the density derived from the transit lightcurve. Because the planet is elongated in the direction of the star, along the line of sight at transit, the planet volume is greater than that inferred from the planet radius derived from transit light curves. This leads to the planet density, and hence the gravity acting on the annulus of atmosphere, being over-estimated. The most strongly affected exoplanets are those which are closest to Roche lobe filling, and tend to be short period. In addition, when taking the depth in full transit, the oblateness of the planet does not need to be considered; however, as discussed in \citet{grant2022}, the oblateness of a planet does affect the shape
of the transit curve at ingress and egress, especially at large impact parameters. Routines to fit the transmission spectrum as a function of $\theta$ (see Fig~\ref{fig:scheme}) have recently been developed \citep{grant2022}, and the oblateness of the planet will need to be considered for this type of analysis.

Accounting for the oblateness of the atmosphere would serve to further decrease the gravity from the value we calculate above. Our quantitative estimates of the magnitude of the effect of rotation are therefore conservative, although we do not expect them to deviate from reality by more than a few ppm.

\subsection{Tidal Locking and Rotation Period}
The type of planets most affected by the centrifugal effect are ultra short period hot Jupiters. We have assumed that these planets are tidally locked, because for these planets, e.g., a Jupiter-like planet orbiting a Sun-like star with a 1 day orbit, the tidal locking timescale ($\sim$10 Myr) is much shorter than the estimated ages of these systems (>1 Gyr).

\section{Conclusion}
In this work, we have analysed the effect of the rotation of an exoplanet on atmospheric retrievals using transmission spectroscopy. We find that for low-density, fast-spinning planets (planets on the lower right section of Figure \ref{fig:boundary}), the centrifugal effect increases the spectroscopic transit depth on the order of 10---100ppm. The combined effect of decreased surface gravity and increased centrifugal acceleration causes the increase in scale height due to rotation to be most prominent on these planets. The difference in the transmission spectrum with and without rotation is more than 30 ppm for the planets labelled with their respective names in the lower right section. Ignoring this effect in our models may lead to retrieved values of gas abundances or temperatures that are significantly different from the true values.
Table~\ref{table:params} demonstrates that the abundances derived for $\textrm{H}_2\textrm{O}$ and $\textrm{CO}_2$ are both displaced from their true values by $\sim 0.3 \textrm{dex}$ (or $\sim 3\sigma$) when rotation is not included in the models used for retrieval.

As the feature amplitudes are increased by the inclusion of the centrifugal forces, the resulting transmission spectrum mimics the properties of one produced by a lighter and hotter atmosphere without centrifugal forces. We also find that the effect, while mostly grey, is a scaled version of the actual transmission spectrum.

We have not included effects such as superrotation or drag of the atmosphere caused by the spinning of the planet. As the quality of data and speed of computational methods improve, more realistic atmospheric models will be required to get a detailed understanding of exoplanet atmospheres.


In conclusion, we recommend that our proposed correction to scale heights, which does not add any significant computational time to the retrievals, should be considered for forward models and atmospheric retrievals of gas-giant exoplanets with orbital periods under 3.0 days.



\section*{Acknowledgements}
AB is supported by a PhD studentship funded by STFC and The Open University. 
JKB is supported by an STFC Ernest Rutherford Fellowship, grant number  ST/T004479/1.
CAH is supported by STFC grants ST/T000295/1 and ST/X001164/1.
SRL is supported by UKSA grants ST/W002949/1 and ST/V005332/1.

We thank the anonymous referee for a very helpful and constructive review.

The NEMESIS code is open-source and can be found at: \url{https://nemesiscode.github.io/}.

The k-tables used for retrievals can be obtained from the Exomol \citep{exomol} database at: \url{https://www.exomol.com/}.

This research has made use of the NASA Exoplanet Archive, which is operated by the California Institute of Technology, under contract with the National Aeronautics and Space Administration under the Exoplanet Exploration Program.

\section*{Data Availability}

The grid of forward models generated for the parameter space exploration (Figure \ref{fig:boundary}) can be found at: \url{https://github.com/riobanerjee/spinny-stuff}.



\bibliographystyle{mnras}
\bibliography{example} 






\bsp	
\label{lastpage}
\end{document}